\documentclass[preprint, traditabstract]{aa} % for a referee version
\usepackage{graphicx}
\usepackage{txfonts}
\def\gtsima{$\;\buildrel > \over \sim \;$}
\def\simgt{\lower.5ex \hbox{\gtsima}}

\begin{document}

   \title{Discovery of interstellar mercapto radicals (SH) \\ with the GREAT instrument on SOFIA}

   \author{D. A. Neufeld\inst{1}, E.~Falgarone \inst{2}, M.~Gerin \inst{2}, B.~Godard\inst{3}, E.~Herbst \inst{4}, G.~Pineau des For\^ets\inst{2,5},  A.~I.~Vasyunin \inst{4}, R.~G\"usten\inst{6}, H.\ Wiesemeyer\inst{6}, and O. Ricken \inst{6,7}}      

   \institute{The Johns Hopkins University, 3400 North Charles St.\, Baltimore, MD 21218, USA \and
LERMA, CNRS UMR 8112, \'Ecole Normale Sup\'erieure \& Observatoire de Paris, Paris, France \and
Departamento de Astrof\'{\i}sica, Centro de Astrobiolog\'{\i}a, CSIC-INTA, Torrej{\'o}n de Ardoz, Madrid, Spain
\and Department of Chemistry, University of Virginia, McCormick Road, P.O. Box 400319,
Charlottesville, VA 22904, USA  \and
Institut d'Astrophysique Spatiale, CNRS UMR 8617, Universit\'e Paris-Sud, Orsay, France \and Max-Planck-Institut f\"ur Radioastronomie, Auf dem H\"ugel 69, 53121 Bonn, Germany
\and I. Physikalisches Institut der Universit\"at zu K\"oln, Z\"ulpicher Strasse 77, 50937 K\"oln, Germany}
 
  \abstract
{We report the discovery of interstellar mercapto radicals along the sight-line to the submillimeter continuum source W49N.  We have used the GREAT instrument on SOFIA to observe the 1383~GHz $^2\Pi_{3/2} \, J = 5/2 \leftarrow 3/2$ lambda doublet in the upper sideband of the L1 receiver.  The resultant spectrum reveals SH absorption in material local to W49N, as well as in foreground gas, unassociated with W49N, that is located along the sight-line.  For the foreground material at velocities in the range 37 -- 44~km/s with respect to the local standard of rest, we infer a total SH column density $\rm \sim 4.6 \times 10^{12}\,\rm cm^{-2}$, corresponding to an abundance of $\rm \sim 7 \times 10^{-9}$ relative to H$_2$, and yielding an SH/H$_2$S abundance ratio $\sim 0.13$.  The observed SH/H$_2$S abundance ratio is much smaller than that predicted by standard models for the production of SH and H$_2$S in turbulent dissipation regions and shocks, and suggests that the endothermic neutral-neutral reaction $\rm SH + H_2 \rightarrow H_2S + H$ must be enhanced along with the ion-neutral reactions believed to produce CH$^+$ and SH$^+$ in diffuse molecular clouds.}

   \keywords{Astrochemistry -- ISM:~molecules -- Submillimeter:~ISM -- Molecular processes -- ISM:~clouds
               }
   \titlerunning{Discovery of interstellar SH}
	\authorrunning{Neufeld et al.}
   \maketitle
%
%________________________________________________________________

\section{Introduction}

In the seven decades following the discovery of interstellar CH (Swings \& Rosenfeld 1937) -- the first identified interstellar molecule -- five or six additional neutral diatomic hydrides have been discovered in the interstellar gas: OH (Weinreb 1963), H$_2$ (Carruthers 1970), HCl (Blake et al.\ 1985), NH (Meyer \& Roth 1991), HF (Neufeld et al.\ 1995), and SiH (Schilke et al.\ 2001; a tentative detection).  These discoveries were obtained from observations over a remarkably wide range of wavelengths, from the far-ultraviolet (H$_2$, discovered at 1013 - 1110~$\AA$), through the near-UV (NH at 3358~$\AA$), the optical (CH at 4300~$\AA$), the far-infrared (HF at 122~$\mu$m), the submillimeter (SiH at 478~$\mu$m, and HCl at 479~$\mu$m), and the radio (OH at 18 cm) spectral regions.  In addition to these neutral molecules, four hydride molecular ions have now been discovered: CH$^+$ (Douglas \& Herzberg 1941), OH$^+$ (Gerin et al.\ 2010a; Wyrowski et al.\ 2010), SH$^+$ (Benz et al.\ 2010), and HCl$^+$ (DeLuca et al.\ 2012; Gupta et al.\ 2012), the latter three in just the past two years.

The diatomic hydrides represent the simplest of interstellar molecules, and -- carefully interpreted in the context of astrochemical models -- may provide key information about the interstellar environment.  For example, the OH$^+$ abundance yields constraints upon the rate of cosmic ray ionization, which initiates the chemical network leading to OH$^+$ (e.g. Neufeld et al.\ 2010).  The CH$^+$, SH$^+$ and OH abundances, by contrast, probe the influence of shocks and turbulent dissipation, which heat the interstellar gas and/or lead to ambipolar diffusion; these processes may thereby drive endothermic chemical reactions that lead to enhanced abundances of CH$^+$, SH$^+$ and OH (e.g.\ Flower, Pineau des For\^ets, \& Hartquist 1985; Godard, Falgarone \& Pineau des For\^ets 2009) 

The mercapto radical, SH -- which, like CH$^+$ and SH$^+$, is expected to show a strong abundance enhancement in warm regions that are heated by shocks or turbulent dissipation -- has been conspicuously absent from the list of previously-detected molecules.  In its ground rotational state, SH can absorb radiation in the $^2\Pi_{3/2}\, J = 5/2 \leftarrow 3/2$ lambda doublet near 1383~GHz, a frequency that is completely inaccessible from the ground, due to atmospheric absorption, and which -- by bad fortune -- falls right in the gap between Bands 5 and 6 of the {\it Herschel Space Observatory}'s Heterodyne Instrument for the Far-Infrared (HIFI).  Fortunately, this spectral region is now accessible to the GREAT (German Receiver for Astronomy at Terahertz Frequencies) instrument\footnote{GREAT is a development by the MPI f\"ür Radioastronomie and the KOSMA / Universit\"at zu K\"oln, in cooperation with the MPI f\"ur Sonnensystemforschung and the DLR Institut f\"ur Planetenforschung.} on SOFIA (the Stratospheric Observatory for Infrared Astronomy), and in this {\it Letter} we report the first detection of interstellar SH with the use of that instrument.  The observations and data reduction are described in \S2, and the results presented in \S3.  In \S4 we discuss the implications of the measured SH column density in the context of astrochemical models. 

%__________________________________________________________________

\section{Observations and data reduction}

We observed the $^2\Pi_{3/2} \, J = 5/2 \leftarrow 3/2$  transition of SH, a lambda doublet for which the strongest hyperfine components ($F=3 \leftarrow 2$) lie at frequencies of 1382.910 and 1383.241 GHz, in the upper sideband of the L1 receiver.  A broad atmospheric ozone feature is present in the lower sideband, at a frequency of 1379.50~GHz.  The LO settings were selected to separate this feature as far as possible (in IF frequency) from the target SH transition. 
%To help determine whether any observed feature did indeed lie in the upper sideband, two observations were carried out with slightly different settings of the local oscillator (LO) frequency.  
These observations, with a combined on-source integration time of 4.0~min, were carried out on 2011 September 28th as part of the  SOFIA Basic Science Program.  The telescope beam, of diameter $\sim 21^{\prime\prime}$ HPBW, was centered on W49N at 
$\rm \alpha=19h\,10m\,13.2s, \delta = +09^0\, 06^\prime \,12.0^{\prime\prime} (J2000)$.  The observations were performed in dual beam switch mode, with a chopper frequency of 1 Hz and the reference positions located 75$^{\prime\prime}$ on either side of the source along an East-West axis. 
The AFFTS backend provided 8192 spectral channels at a spacing of 183.1~kHz.
Thanks to laboratory spectroscopy performed by Morino \& Kawaguchi (1995) and by Klisch et al.\ (1996), the SH rest frequencies are known to estimated accuracy of $<2$~MHz,  The separation of the lambda doublet corresponds to a velocity shift of 71.8~km~s$^{-1}$, and each doublet member is further split into three hyperfine components; the rest frequencies and spontaneous radiative rates are listed in Table 1, for an assumed SH dipole moment of 0.758~D (Meerts \& Dynamus 1975).

\begin{table}
\caption{$^2\Pi_{3/2}\,J = 5/2 \leftarrow 3/2$ SH lines observed with GREAT}
\begin{tabular}{lllr}
\hline
Transition & Frequency & $A_{ij}$ & $\Delta v\,\,\,\,\,\,$ \\
& (GHz) & (s$^{-1}$) & ($\rm km\, s^{-1}$) \\
\hline 
$F=2^+ \leftarrow 2^-$ & 1382.9056 & $0.47 \times 10^{-3}$ & +0.98\\
$F=3^+ \leftarrow 2^-$ & 1382.9101 & $4.72 \times 10^{-3}$ & \\
$F=2^+ \leftarrow 1^-$ & 1382.9168 & $4.24 \times 10^{-3}$ & --1.45\\
$F=2^- \leftarrow 2^+$ & 1383.2365 & $0.47 \times 10^{-3}$ & +1.02\\
$F=3^- \leftarrow 2^+$ & 1383.2412 & $4.72 \times 10^{-3}$ & \\
$F=2^- \leftarrow 1^+$ & 1383.2478 & $4.24 \times 10^{-3}$ & --1.43\\
\hline 
\end{tabular}

\tablefoottext{a}{Velocity shift relative to the strongest hyperfine component of the respective doublet}
\end{table}

The raw data were calibrated to the $T_A^*$ (``forward beam brightness temperature'') scale, fitting independently the dry and the wet content of the atmospheric emission.  Here, the assumed forward efficiency was 0.95 and the assumed beam efficiency for the L1 band was 0.54.  The uncertainty in the flux calibration is estimated to be $\sim 20\%$ (Heyminck et al.\ 2012). 
Additional data reduction -- performed using CLASS\footnote{Continuum and Line Analysis Single-dish Software -- http://www.iram.fr/IRAMFR/GILDAS} -- entailed second-order baseline removal, averaging the data (with a weighting inversely proportional to the square of the r.m.s. noise), and spectral smoothing to a 3.7~MHz channel spacing.

In addition to the SOFIA observations of SH that are the primary subject of this Letter, we have carried out ancillary observations of the $1_{10}-1_{01}$ 168.763~GHz transition of H$_2$S, using the IRAM 30~m telescope located  in Pico Veleta near Granada (Spain) in December 2006 under good weather
conditions. We used the C150 receiver, tuned in single side band, coupled
to two spectrometers: the high resolution correlator VESPA with 40~kHz
spectral resolution, and a broad band filter bank with 1~MHz spectral
resolution. The data were acquired using the wobbler with a frequency of
0.5~Hz, for a total time of 32 minutes, and analyzed with the CLASS
software.

\begin{figure}
%\centering
\includegraphics[width=9.0 cm]{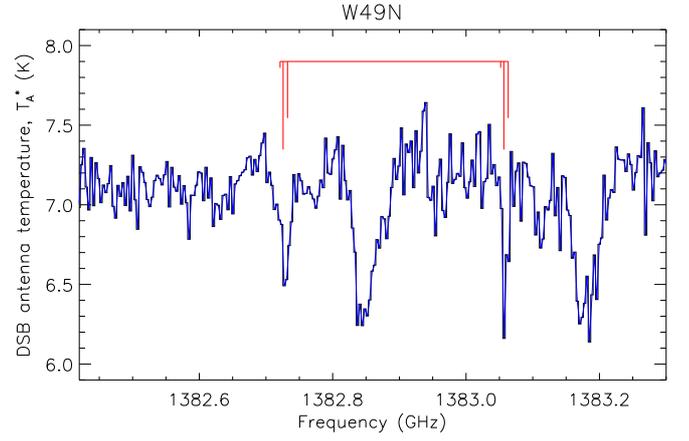}
\caption{Spectrum of SH $^2\Pi_{3/2} \, J=5/2 \leftarrow 3/2$ obtained by GREAT toward W49N. Note that because GREAT employs double sideband receivers, the complete absorption of radiation at a single frequency will reduce the measured antenna temperature to one-half the apparent continuum level.  The lambda doubling and hyperfine splittings are indicated by the red bars for a component at an LSR velocity of 40~$\rm km\,s^{-1}$.
}
\end{figure}

\section{Results}

Figure 1 shows the observed spectrum of SH $^2\Pi_{3/2}\, J=5/2 \leftarrow 3/2$, with the frequency scale corrected to the  Local Standard of Rest (LSR).  The double sideband continuum antenna temperature is $T_A^*({\rm cont}) = 7.2$~K, and the r.m.s noise is 0.12~K in a 3.7~MHz channel.  Because GREAT employs double sideband receivers, the complete absorption of radiation at a single frequency will reduce the measured antenna temperature to one-half the apparent continuum level.

In Figure 2, the fractional transmission is shown separately for each of the lambda doublets (top two panels), with the frequency scale expressed as Doppler velocities relative to the Local Standard of Rest (LSR) for the strongest hyperfine component of each doublet member.  The transmission is given by $2 T_A^* / T_A^*({\rm cont})-1$, given the assumption that the sideband gain ratio is unity.  Analogous spectra are shown for the $1_{10}-1_{01}$ (ground state) transition of ortho-H$_2$S, and for the $^2\Pi_{3/2}\, J=3/2^- \leftarrow\, ^2\Pi_{1/2}\, J= 1/2^+$ transition of CH (Gerin et al.\ 2010b), believed to be a good tracer of H$_2$ (e.g.\ Sheffer et al.\ 2008).

\begin{figure}
%\centering
\includegraphics[width=9.0 cm]{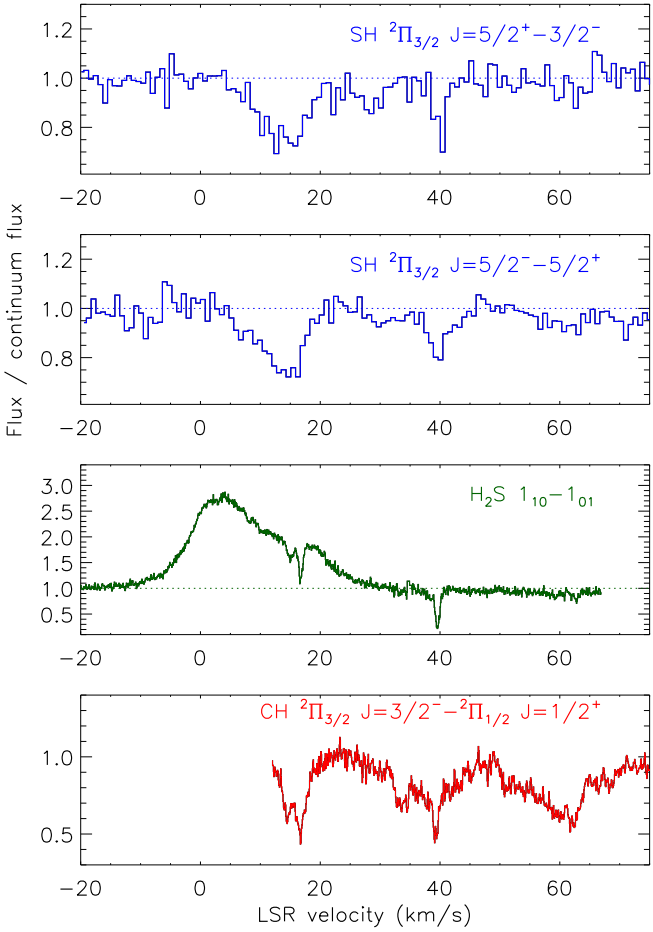}
\vskip 0.2 true in
\caption{Ratio of flux to continuum flux, for SH ($^2\Pi_{3/2}\, J=5/2 \leftarrow 3/2$), ortho-H$_2$S ($1_{10} \leftarrow 1_{01}$), and CH ($^2\Pi_{3/2}\, J=3/2^- \leftarrow \, ^2\Pi_{1/2}\,J=1/2^+$).
The CH spectrum has been hyperfine-deconvolved.}
\end{figure}

Absorption by SH is clearly detected in the range $v_{\rm LSR} \sim 5 - 20\,\rm km\, s^{-1}$, near the systemic velocity of the source.  In addition, a narrow absorption feature is detected unequivocally near $v_{\rm LSR} \sim 39\,\rm km\, s^{-1}$, a component that is clearly present in the spectra of CH, H$_2$S, and many other molecules (e.g.\ Godard et al.\ 2010; Sonnentrucker et al.\ 2010).  This component arises in a foreground cloud unassociated with W49N, which has an (kinematically-) estimated Galactocentric distance of $\sim 6.7$~kpc (Godard et al.\ 2012).  

In contrast to the case of CH, there is an absence of strong SH or H$_2$S absorption in the  $60 - 65\,\rm km\, s^{-1}$ range.  This behavior is similar to that observed for CS (Miyawaki, Hasegawa \& Hayashi 1988) and for the nitrogen hydrides NH, NH$_2$, and NH$_3$ (Persson et al.\ 2012); it may suggest that those molecules -- such as SH -- for which the $v_{\rm LSR} \sim 39\,\rm km\, s^{-1}$ absorption is much stronger than that in the $60 - 65\,\rm km\, s^{-1}$ range all originate in material with a larger molecular fraction.

Because of its large spontaneous radiative decay rate ($4.7 \times 10^{-3}\,\rm s^{-1}$), the SH
transition we have observed possesses a high critical density\footnote{Here, ``critical density'' is defined as the gas density at which the collisional deexcitation rate equals the spontaneous radiative decay rate.} {($\simgt 10^7 \rm cm^{-3}$, for an assumed collisional deexcitation rate $\sim 10^{-10}\rm \, cm^3 s^{-1}$, similar to that computed by Offer et al.\ 1994 for the analogous transition of OH).}  
Thus, in the foreground material unassociated with W49N, we expect that SH will be almost entirely in its ground rotational state, $^2\Pi_{3/2}\,J=3/2$.  In that case, the absorption optical depth, integrated over velocity and summed over the the six components listed in Table 1, is given by
$$\int \tau dv = 2.88 \times 10^{-14} N({\rm SH}) \, \rm cm^2 \, km \, s^{-1}. \eqno(1)$$
(For the 5 - 20 $\rm km\, s^{-1}$ velocity interval, equation (1) would likely yield an underestimate of the SH column density, because -- for gas associated with the dense W49N cloud itself -- there is likely to be a significant SH population in excited rotational states.)

In the present study\footnote{A future study, requiring detailed modeling of the excitation of SH in W49N, will be needed to obtain an estimate of the SH$^+$ column density in the source itself.}, we confine our attention to the narrow absorption feature appearing near $v_{\rm LSR} = 39\rm \, km\, s^{-1}$. 
In Table 2, we present estimates of the SH column density in the $39\rm \, km\, s^{-1}$ absorbing cloud, along with analogous results obtained for ortho-H$_2$S from IRAM 30m observations and for SH$^+$, para-H$_2$O and CH from {\it Herschel}/HIFI observations.  Given an estimated H$_2$ column density of $6.6 \times 10^{20}\rm cm^{-2}$ for this cloud (Godard et al.\ 2012), the observed SH column density implies an SH/H$_2$ abundance ratio $\sim 7 \times 10^{-9}$, corresponding to $\sim 0.003\%$ of the solar abundance of elemental sulfur.  The SH/SH$^+$ ratio is $\sim 1.8$.  For an assumed H$_2$S ortho-to-para of 3 (the value expected in equilibrium), the SH/H$_2$S ratio is 0.13, a value that is signficantly smaller than the $\rm OH/H_2O$ ratio $\sim 1.0$ observed by SOFIA for this 
%same $39\rm \, km\, s^{-1}$ 
absorbing cloud (Wiesemeyer et al.\ 2012)

\begin{table}
\caption{Column densities for $v_{\rm LSR}$ in the 37 -- 44 $\rm km\,s^{-1}$ range}
\begin{tabular}{lccl}
\hline
Molecule & Column  & Abundance$^a$\\
& density & relative to H$_2$ \\
& (cm$^{-2}$) \\
\hline 
SH 		& $4.6 \times 10^{12}$   & $6.9 \times 10^{-9}$ & Present work\\
ortho-H$_2$S 	& $2.6 \times 10^{13}$   & $3.9 \times 10^{-8}$ & Present work \\
SH$^+$ 		& $2.6 \times 10^{12}$   & $3.9 \times 10^{-9}$ & Godard et al.\ 2012\\
para-H$_2$O 	& $1.5\times 10^{13}$    & $2.3 \times 10^{-8}$ & Sonnentrucker et al.\ 2010\\
CH		& $5.8 \times 10^{13}$   & $9.0   \times 10^{-8}$ & Gerin et al.\ 2012\\
%HF 		& $> 2.4\times 10^{13}$   $>3.6 \times 10^{-8}$ & Godard et al.\ 2012\\

\hline 
\end{tabular}
\tablefoottext{a}{For an assumed H$_2$ column of $6.6 \times 10^{20} \rm \, cm^{-2}$ (Gerin et al.\ 2012)}
\end{table}

\section{Discussion}

Sulfur is unusual among the abundant elements in having a set of hydrides and hydride cations with relatively small bond energies.  Thus {\it none} of the species S, SH, S$^+$, SH$^+$, or H$_2$S$^+$ can undergo an exothermic H atom abstraction reaction with H$_2$. (In other words, $\rm X + H_2 \rightarrow XH + H$ is endothermic for X = S, SH, S$^+$, SH$^+$, or H$_2$S$^+$.)  Figure 3 illustrates the thermochemistry of the sulphur bearing hydrides by displaying the heats of formation of various sulphur-bearing species (with associated hydrogen atoms and H$_2$ molecules involved in their formation.)  The endothermicities of the reactions $\rm X + H_2 \rightarrow XH + H$, expressed in temperature units, $\Delta E/k$, are 9641, 6984, 10117, 7634, and 4096 K respectively for X = S, SH, S$^+$, SH$^+$, and H$_2$S$^+$.

\begin{figure}
%\centering
\includegraphics[width=8.4 cm]{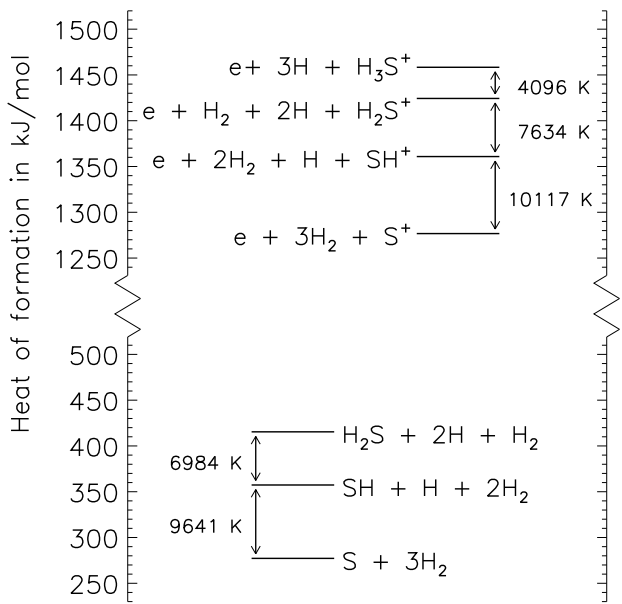}
\vskip 0.2 true in
\caption{Heats of formation of various S-bearing species (with associated H atoms produced during their formation), based upon thermochemical data from the NIST Chemistry Web book.}
\end{figure}

Clearly, these H atom abstraction reactions leading to the formation of the sulfur-containing hydrides 
are negligibly slow at the temperatures ($< 100$~K) typical of diffuse or dense molecular clouds.  Nevertheless, surprisingly high column densities of H$_2$S have previously been observed in both diffuse molecular clouds and dense regions of active star-formation.  For example, based on observations of diffuse foreground material along the sight-lines to Sgr B2 (M) and W49N, Tieftrunk et al. (1994) measured $\rm H_2S$ abundances (of several $\times 10^{-9}$ relative to H$_2$ that were considerably larger than those predicted in standard chemical models,  They suggested that $\rm H_2S$ might have been enhanced by high temperature reactions in shocks (e.g. Pineau des For\^ets, Roueff, \& Flower 1986) or by hydrogenation reactions on grain surfaces.  Some smaller $\rm H_2S$ abundances (of several times $10^{-10}$) were observed in diffuse clouds at high Galactic latitude by Lucas \& Liszt (2002), while even larger $\rm H_2S$ abundances (up to $\sim 10^{-6}$) have been inferred for dense regions of active star-formation (Minh et al. 1991).  

More recently, enhanced SH$^+$ abundances $\sim \rm few \times 10^{-9}$ have been measured in the diffuse ISM, using APEX (Menten et al.\ 2011) and {\it Herschel}/HIFI (Godard et al.\ 2012).  These measurements have been interpreted as resulting from an increased reaction rate for $\rm S^+ + H_2 \rightarrow SH^+ + H$ in turbulent dissipation regions where the gas temperature is elevated and significant ion-neutral drift is present.   The small SH/H$_2$S  abundance ratio $\sim 0.13$ inferred in \S3 above for the $39\rm \, km\, s^{-1}$ absorbing cloud suggests that the reaction $\rm SH + H_2 \rightarrow H_2S + H$ must be similarly increased, and argues that endothermic neutral-neutral reactions are enhanced along with ion-neutral reactions.  As described below, this conclusion is supported by detailed modeling of (1) standard photodissociation regions (PDRs); (2) turbulent dissipation regions (TDRs); and (3) 'C'- and 'J'-type shock waves.

With the use of the Meudon PDR model (Le~Petit et al.\ 2006), we have found that standard PDR models -- i.e. models that do not include ion-neutral drift or additional heating mechanisms beyond those associated with ultraviolet irradiation and cosmic rays -- greatly overpredict the $\rm SH/H_2S$ ratio in low density diffuse gas, yielding typical values of $10^4$, and greatly underpredict the $\rm SH/H_2$ and $\rm H_2S/H_2$ ratios.  We found that $\rm SH/H_2S$ ratios as small as unity are predicted only in gas clouds with densities, $n_{\rm H}$, larger than $10^4$~cm$^{-3}$, and visual extinctions, $A_{\rm V}$ larger than 4, values that are both unreasonably large for the $39\rm \, km\, s^{-1}$ absorbing cloud toward W49N.  We have also investigated PDR models that include the effects of grain surface hydrogenation of S and SH and the subsequent photodesorption of H$_2$S (Vasyunin \& Herbst 2011); these too underpredict the $\rm SH/H_2$ and $\rm H_2S/H_2$ ratios, unless an unreasonable large gas density is assumed.  Using models for the chemistry of sulfur-bearing molecules in TDRs (Godard et al.\ 2009) and 'C'- and 'J'-type shocks (Pineau des For\^ets et al.\ 1986), we determined that the predicted $\rm SH/H_2S$ ratios in diffuse molecular clouds are $\sim 10$, significantly smaller than those for standard PDRs models without grain surface reactions, but nevertheless a factor $\sim 100$ larger than what is observed.  Here, the production of SH is enhanced by the sequence $\rm S^+(H_2, H)SH^+(H_2, H)H_2S^+(e,H)SH$, in which the first two endothermic reactions are significantly enhanced by ion-neutral drift.  While the predicted SH and SH$^+$ abundances are broadly consistent with the observations, H$_2$S is underpredicted by a factor of 100.  As in the PDR models, smaller SH/H$_2$S abundances are only predicted in clouds with implausibly  large $n_{\rm H}$ and A$_{\rm V}$.  Apparently, standard models for PDRs, TDRs, or 'C'- or 'J'-type shocks cannot account for the observed SH/H$_2$S ratio, because they fail to predict a sufficient enhancement in the rates of neutral-neutral reactions.  

We note however that existing models for sulfur chemistry in TDRs and 'C'-type shocks assume that all neutral species share a common velocity.  As originally pointed out by Flower \& Pineau des For\^ets (1998) in the context of CH and CH$^+$ chemistry, neutral molecules that are produced by dissociative recombination of molecular ions in environments where there is significant ion-neutral drift -- such as S and SH in the case of present interest -- initially carry an imprint of the velocity of the ionized parents from which they formed.  This effect may enhance endothermic reactions with H$_2$ that occur before the newly-formed neutral species has undergone sufficient elastic collisions to acquire the velocity of the neutral fluid.  Detailed modeling, which we defer to a future publication, will be needed to determine whether the SH/H$_2$S ratio serves as a diagnostic of this effect.

\begin{acknowledgements}
Based on observations made with the NASA/DLR Stratospheric Observatory for Infrared Astronomy. SOFIA Science Mission Operations are conducted jointly by the Universities Space Research Association, Inc., under NASA contract NAS2-97001, and the Deutsches SOFIA Institut under DLR contract 50 OK 0901.  This research was supported by USRA through a grant for Basic Science Program 81-0014. EF, MG and BG acknowledge support from the French CNRS/INSU Programme
PCMI (Physique et Chimie du Milieu Interstellaire).
\end{acknowledgements}


\begin{thebibliography}{}

\bibitem[Asplund et al.(2009)]{As} Asplund, M., Grevesse, N., Sauval, A.~J., Scott, P. 2009,
ARA\&A, 47, 481.
\bibitem[Benz et 
al.(2010)]{2010A&A...521L..35B} Benz, A.~O., Bruderer, S., van Dishoeck, E.~F., et al.\ 2010, \aap, 521, L35 
\bibitem[]{}Blake, G. A., Keene, J., \& Phillips, T. G. 1985, ApJ, 295, 501 
\bibitem[Carruthers(1970)]{1970ApJ...161L..81C} Carruthers, G.~R.\ 1970, 
\apjl, 161, L81 
\bibitem[]{}Douglas, A. E., \& Herzberg, G. 1941, ApJ, 94, 381 
\bibitem[]{}DeLuca et al.\ 2012, ApJ, submitted
\bibitem[Flower et al.(1985)]{1985MNRAS.216..775F} Flower, D.~R., Pineau 
des For\^ets, G., \& Hartquist, T.~W.\ 1985, \mnras, 216, 775 
\bibitem[]{}Flower, D. R.,  \& Pineau des For\^ets, G.  1998, MNRAS, 297, 1182
\bibitem[Gerin et 
al.(2010)]{2010A&A...518L.110G} Gerin, M., de Luca, M., Black, J., et al.\ 2010a, \aap, 518, L110 
\bibitem[Gerin et 
al.(2010)]{2010A&A...521L..16G} Gerin, M., de Luca, M., Goicoechea, J.~R., et al.\ 2010b, \aap, 521, L16 
\bibitem[]{}Gerin, M.\ et al.\ 2012, A\&A, in preparation
\bibitem[]{}Godard, B., Falgarone, E., Gerin, M., et al.\ 2012, A\&A, in press 
\bibitem[]{}Godard, B., Falgarone, E., Gerin, M., Hily-Blant, P., \& De Luca, M. 2010, A\&A, 520, A20+
\bibitem[]{}Godard, B., Falgarone, E.,  \& Pineau Des For\^ets, G. 2009, A\&A, 495, 847 
\bibitem[]{}Gupta, H.\ et al.\ 2012, ApJ, in preparation
\bibitem[]{}Heyminck, S., Graf, U.~U., G\"usten, R.\ et al.\ 2012, A\&A, this volume
\bibitem[]{}Klisch, E., Klaus, T., Belov, S. P., et al.\ 1996,       ApJ, 473, 1118
\bibitem[Le Petit et al.(2006)]{2006ApJS..164..506L} Le Petit, F., 
Nehm{\'e}, C., Le Bourlot, J., \& Roueff, E.\ 2006, \apjs, 164, 506 
\bibitem[]{}Lucas, R., \& Liszt, H. S. 2002, A\&A, 384, 1054
\bibitem[]{}Meerts, W. L., \& Dymanus, A.\ 1975, Can J. Phys. 53, 2123
\bibitem[]{}Menten, K. M., Wyrowski, F., Belloche, A., et al.\ 2011, A\&A, 525, A77+
\bibitem[]{}Meyer, D. M., \& Roth, K. C.  1991, ApJ, 376, L49 
\bibitem[Miyawaki et al.(1988)]{1988PASJ...40...69M} Miyawaki, R., 
Hasegawa, T., \& Hayashi, M.\ 1988, \pasj, 40, 69 
\bibitem[Minh et al.(1991)]{1991ApJ...366..192M} Minh, Y.~C., Ziurys, 
L.~M., Irvine, W.~M., \& McGonagle, D.\ 1991, \apj, 366, 192 
\bibitem[]{}Morino, I., and Kawaguchi, K. 1995, J.\ Mol.\ Spectrosc.\ , 170, 172 
\bibitem[]{}Neufeld, D. A., Zmuidzinas, J., Schilke, P., \& Phillips, T. G. 1997, ApJ, 488, L141 
\bibitem[Neufeld et 
al.(2010)]{2010A&A...521L..10N} Neufeld, D.~A., Goicoechea, J.~R., Sonnentrucker, P., et al.\ 2010, \aap, 521, L10 
\bibitem[Offer et al.(1994)]{1994JChPh.100..362O} {Offer, A.~R., van Hemert, 
M.~C., \& van Dishoeck, E.~F.\ 1994, \jcp, 100, 362 }
\bibitem[]{}Pineau des For\^ets, G., Roueff, E., \& Flower, D. R. 1986, MNRAS, 223, 743
\bibitem[]{}Schilke, P., Benford, D. J., Hunter, T. R., Lis, D. C., \& Phillips, T. G. 2001, ApJS, 132, 281 
\bibitem[Sheffer et al.(2008)]{2008ApJ...687.1075S} Sheffer, Y., Rogers, 
M., Federman, S.~R., et al.\ 2008, \apj, 687, 1075 
\bibitem[Sonnentrucker et 
al.(2010)]{2010A&A...521L..12S} Sonnentrucker, P., Neufeld, D.~A., Phillips, T.~G., et al.\ 2010, \aap, 521, L12 
\bibitem[]{}Swings, P., \& Rosenfeld, L. 1937, ApJ, 86, 483
\bibitem[]{}Tieftrunk, A., Pineau des For\^ets, G., Schilke, P., \& Walmsley, C. M. 1994, A\& A, 289, 579 
\bibitem[Vasyunin 
\& Herbst(2011)]{2011IAUS..280E..29V} Vasyunin, A.~I., \& Herbst, E.\ 2011, in The Molecular Universe, Proceedings of the 280th Symposium of the International Astronomical Union held in Toledo, Spain, May 30-June 3, 2011
\bibitem[]{}Weinreb, S. 1963, Nature, 200, 829 
\bibitem[]{}Wiesemeyer, H.\ et al.\ 2012, A\&A, this volume

\bibitem[Wyrowski et 
al.(2010)]{2010A&A...518A..26W} Wyrowski, F., Menten, K.~M., G{\"u}sten, R., \& Belloche, A.\ 2010, \aap, 518, A26 



\end{thebibliography}
\end{document}